\documentclass[preprint]{revtex4}

\usepackage{graphicx}
\usepackage{mathptmx}
\usepackage{flushend}
\usepackage[colorlinks,citecolor=blue,urlcolor=blue,linkcolor=blue]{hyperref}

\usepackage{hypernat}

\usepackage{bm}
\usepackage{amssymb}
\usepackage{amsmath}

\usepackage{slashed}

\usepackage{graphics}
\usepackage[export]{adjustbox}

\usepackage{xparse}

\usepackage{orcidlink}

\usepackage[caption=false,labelfont=normalsize,labelformat=empty,textfont=normalsize,justification=centering]{subfig}

\renewcommand{\Im}{\mathrm{Im}}
\newcommand{\D}{\mathrm{d}}
\newcommand{\iu}{\mathrm{i}}
\newcommand{\eq}{\mathrm{eq}}
\newcommand{\TR}{T_{\mathrm{R}}}
\newcommand{\lbar}{\Bar{l}}
\newcommand{\Hbar}{\Bar{H\vphantom{l}\hskip0.35mm}\hskip-0.35mm}

\begin{document}

\title{Minimal effective theory for leptogenesis, dark matter, and neutrino masses}

\author{Tom\'a\v s Bla\v zek\,\orcidlink{0000-0002-2645-0283}}
\email{tomas.blazek@fmph.uniba.sk}
\author{Peter Mat\'ak\,\orcidlink{0000-0003-1420-1148}}
\email{peter.matak@fmph.uniba.sk}
\author{J\'an Ramaj\,\orcidlink{0009-0002-7420-6709}}
\author{Martina Sabov\'a\,\orcidlink{0009-0009-0759-4036}}

\affiliation{Department of Theoretical Physics, Faculty of Mathematics, Physics and Informatics, Comenius University in Bratislava, Mlynsk\'a dolina, 84248 Bratislava, Slovak Republic}

\date{\today}

\begin{abstract}
We study the phenomenology of choosing a minimal set of effective operators simultaneously generating the dark matter relic density and matter-antimatter asymmetry of the universe. Neutrino masses are obtained in a specific case of baryogenesis via leptogenesis. We find that only two new particles---a heavy unstable fermion and a light dark matter scalar---need to be included in addition to the Standard Model particle content.  
\end{abstract}

\maketitle

\section{Introduction}\label{sect1}

The nature of the invisible matter affecting the galaxy dynamics and the universe's expansion, along with the unknown origin of the matter-antimatter asymmetry, represent two unresolved questions at the crossroads of contemporary particle physics and cosmology. Let us assume that both should be explained within a simple extension of the Standard Model. How much complexity in terms of particles and their interactions is necessary? What can the minimal model look like? In the literature, several ap\-proach\-es can be recognized based on whether the dark matter relic density freezes in \cite{Bento:2001rc, Bhattacharya:2021jli, Goudelis:2021qla, Datta:2021elq, Hall:2010jx, Biswas:2018sib, Goudelis:2022bls, Barman:2021tgt} or freezes out \cite{Davoudiasl:2010am, Baldes:2014gca, Baldes:2014rda, Ciscar-Monsalvatje:2023zkk, Falkowski:2011xh, Chun:2011cc, Cui:2011ab, Cui:2012jh, Baldes:2015lka, Chu:2021qwk, Heisig:2024mwr} and whether the model is renormalizable \cite{Bento:2001rc, Falkowski:2011xh, Chun:2011cc, Cui:2011ab, Cui:2012jh, Baldes:2015lka, Bhattacharya:2021jli, Goudelis:2021qla, Chu:2021qwk, Datta:2021elq, Heisig:2024mwr, Maleknejad:2020pec} or relies on high\-er-di\-men\-sion\-al effective operators \cite{Davoudiasl:2010am, Baldes:2014gca, Baldes:2014rda, Goudelis:2022bls, Barman:2021tgt, Ciscar-Monsalvatje:2023zkk}. Unlike the previous studies, here we construct the effective description from the bottom up, following the minimality principle and including only the essential ingredients needed to generate the matter-antimatter asymmetry while the lightest new particle remains stable. Remarkably, we achieve this with only two new particles interacting with the Standard Model through a single effective portal operator. 

The paper is structured as follows. In Sect.~\ref{sect2}, we discuss general arguments that lead to an effective theory in which $CP$-violating reactions produce a baryon or lepton number asymmetry. In Sect.~\ref{sect3}, the minimal version of the model based on leptogenesis via ultraviolet freeze-in is presented within a simple analytical approximation. Numerical solutions of the full set of the Boltzmann equations are studied in Sect.~\ref{sect4}, where we also consider a freeze-out scenario. Finally, we summarize in Sect.~\ref{sect5}.

\section{{\it CP} asymmetries with effective operators}\label{sect2}

\begin{figure}[t!]
\subfloat{\label{fig1a}}
\subfloat{\label{fig1b}}
\subfloat{\label{fig1c}}
\centering\includegraphics[scale=1]{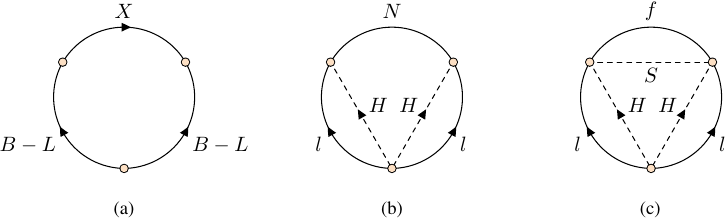}
{\caption{\label{fig1}\raggedright 
Vacuum diagram made of the vertices corresponding to the operators in Eq.~\eqref{eq3}, where each line represents all of Wick's contractions between two operators ({\bf a}). Vac\-u\-um diagram for the invariants of effective Lagrangians in Eq.~\eqref{eq4} ({\bf b}) and Eq.~\eqref{eq6} ({\bf c}), respectively.}}
\end{figure}

According to Sakharov's conditions \cite{Sakharov:1967dj}, out-of-equi\-lib\-ri\-um $CP$-violating processes are necessary to generate a mat\-ter-antimatter asymmetry from an initially symmetric state.  Let us, therefore, look closely at how the $CP$ asymmetries occur. We denote $S_{fi}=\delta_{fi}+\iu T_{fi}$ the scattering matrix elements for the $i\rightarrow f$ reaction, defining the $CP$ asymmetry
\begin{align}\label{eq1}
\Delta\vert T_{fi\vphantom{\Bar{f}}}\vert^2=\vert T_{fi\vphantom{\Bar{f}}}\vert^2-\vert T_{\Bar{f}\Bar{i}}\vert^2
=\vert T_{fi\vphantom{\Bar{f}}}\vert^2-\vert T_{if\vphantom{\Bar{f}}}\vert^2
\end{align}
with the last equality resulting from the $CPT$ invariance. The unitarity condition $S^\dagger S=1$ implies $\iu T^\dagger=\iu T-\iu T^\dagger\iu T$ that can be iteratively used in Eq.~\eqref{eq1} obtaining \cite{Blazek:2021olf}
\begin{align}\label{eq2}
\Delta \vert T^{\vphantom{\dagger}}_{fi}\vert^2=&\sum_{n}(\iu T_{in} \iu T_{nf} \iu T_{fi} - \iu T_{if} \iu T_{fn} \iu T_{ni})\\
&-\sum_{n,m}(\iu T_{in} \iu T_{nm} \iu T_{mf} \iu T_{fi} - \iu T_{if} \iu T_{fm} \iu T_{mn} \iu T_{ni})\nonumber\\
&+\ldots\nonumber
\end{align}
where  only the first row contributes at the lowest order. For a specific set of particles in the intermediate state $n$, the two terms can be understood as forward-scattering diagrams with at least two holomorphic cuts \cite{Coster:1970jy, Bourjaily:2020wvq, Hannesdottir:2022bmo}. Joining together the external legs of each diagram produces two cyclic or vacuum diagrams that only differ by the direction of arrows on propagators \cite{Roulet:1997xa, Blazek:2021olf}. Vacuum diagrams can also be associated with basis invariants \cite{Botella:2004ks, Nilles:2018wex, Bento:2023owf}. These are $CP$-odd, i.e. come with an irreducible complex phase, if the respective vacuum diagrams change under the reversal of arrows on their internal lines, which is also necessary for an asymmetry to occur from Eq.~\eqref{eq2}.

What is the minimum set of vertices allowing for a matter asymmetry generation? Starting with a vacuum diagram, at least one of its internal lines must correspond to an out-of-equilibrium particle, and at least two vertices must violate the $B-L$ number \cite{Sakharov:1967dj}. Let us then consider an operator $\hat{O}_{B-L}$ that is a singlet with respect to the Standard-Model gauge group and carries a $B-L$ number. For instance, it might be a product of the lepton and Higgs doublets or a colorless combination of three quark fields. Note that it does not have to be Lorentz invariant. In the Lagrangian density, we couple $\hat{O}_{B-L}$ to another operator $\hat{O}_X$ and, if not the same, to charge-conjugated $\hat{O}^c_X$. Rephasing the fields in $\hat{O}_{B-L}$ and $\hat{O}_X$ can make the respective couplings real. But we can also include the square of $\hat{O}_{B-L}$ inevitably coming with an irreducibly complex coupling. That itself would not be all that interesting if it were not possible to combine the matrix elements of the effective operators
\begin{align}\label{eq3}
\hat{O}^{\vphantom{\dagger}}_{B-L}\hat{O}^{\vphantom{\dagger}}_X, \hat{O}^{\vphantom{\dagger}}_{B-L}\hat{O}^{c\vphantom{\dagger}}_X, 
\hat{O}^{\dagger 2}_{B-L}
\end{align}
into a vacuum diagram that can be cut as in Eq.~\eqref{eq2}. However, by construction, we can always draw the diagram in Fig.~\ref{fig1a}. In this diagram, the $B-L$ number is violated in all three vertices. Therefore, if $\hat{O}_X$ is made of fields from which at least one represents an out-of-equi\-lib\-ri\-um particle, cutting the diagram in Fig.~\ref{fig1a} leads to reaction asymmetries sourcing the $B-L$ number in the Boltzmann equation.

To make the general arguments more understandable, we use a straightforward example that might be familiar to many readers. Let us put $\hat{O}^{\vphantom{c}}_{B-L}=lH$, where $l$ and $H$ denote the lepton and Higgs doublets, respectively, and let $\hat{O}^{\vphantom{c}}_X=\hat{O}^c_X=N$ be a massive Majorana fermion. We obtain the effective Lagrangian density
\begin{align}\label{eq4}
L_{\text{eff}}=\lambda\Bar{N}P_L lH+\frac{\lambda'}{\Lambda}H\lbar ^c P_L lH
+\text{H.c.}
\end{align}
where flavour and isospin indices are implicit. The model in Eq.~\eqref{eq4}
corresponds to the seesaw type-I leptogenesis scenario with $N=N_1$ \cite{Fukugita:1986hr}. The heavier Majorana neutrinos are integrated out, resulting in the second term in Eq.~\eqref{eq4}. We can cut the $N$-line in Fig.~\ref{fig1b}, which gives us a two-loop self-energy diagram that can be further cut twice. Subtracting the result from a similarly cut diagram with reversed propagator arrows, i.e. conjugated couplings, we obtain the asymmetry in the $N\rightarrow lH$ squared amplitude proportional to $\Im[\lambda^2\lambda'^*]$. To this perturbative order, all the asymmetries of the model result from Fig.~\ref{fig1b} and its conjugation.\footnote{See Refs.~\cite{Blazek:2021olf, Botella:2004ks} for more details.}

The advantage of the model in Eq.~\eqref{eq4} is the ability to explain the matter-antimatter asymmetry and the tiny neutrino masses. However, it does not offer a stable dark matter candidate. In the next section, we modify the model to find an effective solution to all three problems.

\section{The minimal model}\label{sect3}

\begin{figure*}
\subfloat{\label{fig2a}}
\subfloat{\label{fig2b}}
\subfloat{\label{fig2c}}
\subfloat{\label{fig2d}}
\subfloat{\label{fig2e}}
\subfloat{\label{fig2f}}
\centering\includegraphics[scale=1]{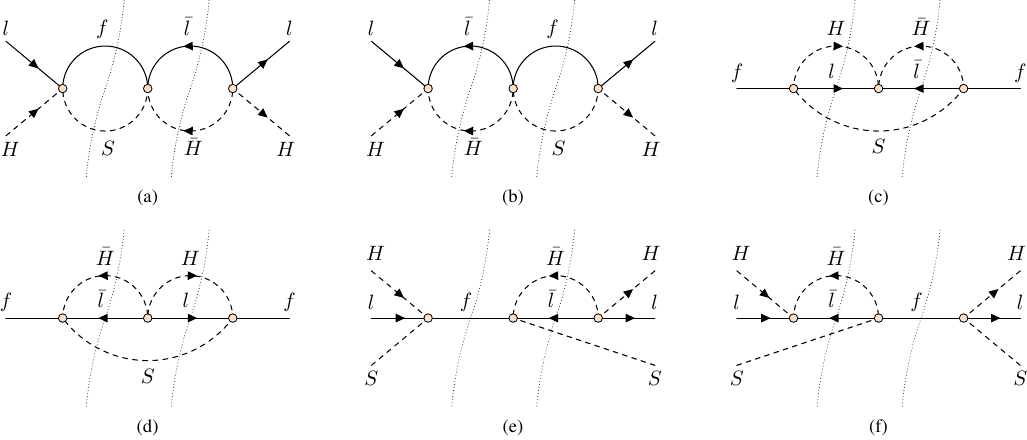}
\caption{\label{fig2} Forward-scattering diagrams obtained from Fig.~\ref{fig1c} contributing to the asymmetry of $lH\rightarrow Sf$, $lH\rightarrow\lbar\Hbar$ ({\bf a}, {\bf b}), $f\rightarrow lHS$, $f\rightarrow\lbar\Hbar S$ ({\bf c}, {\bf d}), and $lHS\rightarrow f$, $lHS\rightarrow\lbar\Hbar S$ ({\bf e}, {\bf f}) according to Eq.~\eqref{eq2}. Each cut represents integration over the on-shell momenta and summation over the spins \cite{Blazek:2021olf}.}
\end{figure*}
To make the lightest dark-sector particle stable, we assume it is charged under a symmetry group $G$. The operator $\hat{O}_X$ couples to the Standard Model; therefore, even if made of dark-sector fields, $\hat{O}_X$ must be a singlet with respect to $G$. Finally, with $\hat{O}^{\vphantom{c}}_{B-L}=lH$, operator $\hat{O}_X$ transforms as a spinor with respect to the Lorentz group. 

For a single-component dark sector, the most straightforward option corresponds to $\hat{O}_X$ being a product of three fermionic $\chi$ fields charged under the $\mathbb{Z}(3)$ symmetry and coupled to the Standard Model as
\begin{align}\label{eq5}
\frac{\lambda_1}{\Lambda^3}(\Bar{\chi}^cP_R\chi)(\Bar{\chi}^cP_L l)H
+\frac{\lambda_2}{\Lambda^3}(\Bar{\chi}P_R\chi^c)(\Bar{\chi}P_L l)H
+\text{H.c.}
\end{align}
Unfortunately, in this scenario, the matter-antimatter asymmetry would be very suppressed due to the small value of the $\hat{O}^2_{B-L}$ coupling constrained by neutrino masses. To overcome the suppression, we propose $\hat{O}^{\vphantom{c}}_X=\hat{O}^c_X=Sf$, where $f$ is a Majorana fermion of mass $m$ and $S$ is a real scalar of mass $m'$. Both $f$ and $S$ carry a negative dark-sector parity; hence, $G=\mathbb{Z}(2)$. The effective Lagrangian then reads
\begin{align}\label{eq6}
L_{\text{eff}}=\frac{\lambda}{\Lambda} S\Bar{f}P_L lH+\frac{\lambda'}{\Lambda}H\lbar ^c P_L lH+\text{H.c.}
\end{align}
and, as we will see below, allows for the simultaneous generation of the matter-antimatter asymmetry, dark matter relic density, and neutrino masses. Before analyzing several benchmark scenarios, let us look at the $CP$ asymmetries entering the calculation.

\subsection{$CP$ asymmetries and unitarity constraints}

Following the procedure discussed in the previous section, we cut the vacuum diagram in Fig.~\ref{fig1c} and its conjugation into three pieces in all possible ways. The resulting asymmetry contributions are shown in Fig.~\ref{fig2}. For the $lH\rightarrow Sf$ reaction, we introduce
\begin{align}\label{eq7}
\Delta\gamma^\eq_{lH\vphantom{\lbar}\rightarrow Sf}=\gamma^\eq_{lH\vphantom{\lbar}\rightarrow Sf}-\gamma^\eq_{\lbar\Hbar \rightarrow Sf}
\end{align}
where
\begin{align}\label{eq8}
\gamma^\eq_{i\rightarrow f}=
\int\prod_{\forall i}[\D\bm{p}_i]\exp\{-E_i/T\}\int\prod_{\forall f}[\D\bm{p}_f]\frac{1}{V_4}
\vert T_{fi}\vert^2   
\end{align}
is the equilibrium rate of the $i\rightarrow f$ process, $V_4$ denotes a four-dimensional volume, and $[\D\bm{p}]=\D^3\bm{p}/(2\pi)^3/(2E)$. The asymmetry in Eq.~\eqref{eq7} results from the difference of the diagrams in Figs.~\ref{fig2a} and \ref{fig2b} integrated with the Maxwell-Boltz\-mann densities. The $lH\rightarrow\lbar\Hbar$ asymmetry receives contributions from the same diagrams subtracted in a reversed order. Therefore, the two asymmetries only differ in sign, and in agreement with the $CPT$ and unitarity constraints \cite{Kolb:1979qa, Dolgov:1979mz}
\begin{align}\label{eq9}
\Delta\gamma^\eq_{lH\vphantom{\lbar}\rightarrow Sf}+\Delta\gamma^\eq_{lH\rightarrow\lbar\Hbar}=0.    
\end{align}

We can similarly obtain the asymmetry of the $f\rightarrow lHS$ decay from Figs.~\ref{fig2c} and \ref{fig2d}. The inverse decays, however, imply the inclusion of the $lHS\rightarrow\lbar\Hbar S$ reaction, in which the state of the $S$ particle remains unchanged. The respective asymmetry is obtained by subtracting the crocodilian diagram \cite{Hochberg:2022jfs} in Fig.~\ref{fig2e} from Fig.~\ref{fig2f}, and 
\begin{align}\label{eq10}
\Delta\gamma^\eq_{lHS\rightarrow\lbar\Hbar S\vphantom{f}}+\Delta\gamma^\eq_{lHS\rightarrow f\vphantom{\lbar}}=0.    
\end{align}
Note that omitting the $lHS\rightarrow\lbar\Hbar S$ peculiar reaction results in the violation of Sakharov's conditions \cite{Sakharov:1967dj}, as the lepton number will be produced in the inverse decays even in thermal equilibrium.

\subsection{Approximate analytical solutions for leptogenesis via $f$ and $S$ freeze-in}

Assuming negligible $S$ and $f$ initial abundances at the reheating temperature $\TR$, these particles are produced via ultraviolet freeze-in \cite{Elahi:2014fsa}. If the rates of the two contributing reactions, $lH\rightarrow Sf$ and $\lbar\Hbar \rightarrow Sf$, are not the same, a lepton asymmetry is produced. In other words, the asymmetry of the $lH\rightarrow Sf$ reaction sources the lepton number. 

For simplicity, let us first assume that dark-sector masses are much smaller than 
$\TR$, allowing us to solve the Boltzmann equations analytically. The reaction rates have a mass dimension of four, while the asymmetries are proportional to the third inverse power of the energy scale $\Lambda$. If the masses were neglected entirely, obtaining an odd dimension from fermionic traces would not be possible, and the asymmetry would vanish. Therefore, we begin with including one power of the fermionic mass $m$ in the asymmetry and neglecting the masses elsewhere, obtaining
\begin{align}\label{eq11}
\Delta\gamma^\eq_{lH\vphantom{\lbar}\rightarrow Sf}=\frac{3\Im[\lambda^2\lambda'^*]}{32\pi^6}\frac{mT^6}{\Lambda^3}=s(\TR)H(\TR)\times Ax^{-6}
\end{align}
for the asymmetry, and
\begin{align}\label{eq12}
\gamma^\eq_{lH\vphantom{\lbar}\rightarrow Sf}+\gamma^\eq_{\lbar\Hbar \rightarrow Sf}=
\frac{\vert\lambda\vert^2}{8\pi^5}\frac{T^6}{\Lambda^2}=s(\TR)H(\TR)\times Sx^{-6}
\end{align}
for the leading-order symmetric part. Here $s(T)$ and $H(T)$ are the entropy density and the Hubble parameter at the temperature $T$, respectively, while $x=\TR/T$. We also used an abbreviated notation
\begin{align}\label{eq13}
\Im[\lambda^2\lambda'^*]=\sum_{a,b}\Im[\lambda^{\vphantom{*}}_a\lambda'^*_{ab}\lambda^{\vphantom{*}}_b], \vert\lambda\vert^2=\sum_a\vert\lambda_a\vert^2
\end{align}
with sums over the lepton flavour indices $a,b$. Finally, $\lambda'$ can be related to neutrino masses as
\begin{align}\label{eq14}
m_\nu=\sqrt{\sum_{a\vphantom{b}} m^2_{\nu_a}}=\frac{v^2}{\Lambda}\vert\lambda'\vert\quad\text{with}\quad
\vert\lambda'\vert^2=\sum_{a,b}\vert\lambda'_{ab}\vert^2
\end{align}
where $v=246~\text{GeV}$ is the vacuum expectation value of the Higgs field after the electroweak symmetry breaking. We approximate $m^2_\nu$ by $\vert\Delta m^2_{32}\vert\simeq 2.5\times 10^{-3}~\text{eV}^2$ \cite{ParticleDataGroup:2024cfk, Simkovic:2021tkr, Simkovic:2025} in the rest of this work. 

For fermions $f$, and similarly for other particles, let us denote $Y_f$ the number density divided by $s$. In the freeze-in case, neglecting the washout effects may lead to a reasonable approximation. The Boltzmann equations for $Y_f$, $Y_S$, and the lepton number asymmetry $Y_{\Delta L}$ then take a simple form of
\begin{align}\label{eq15}
\frac{\D Y_f}{\D x}=\frac{\D Y_S}{\D x}=Sx^{-2}, 
\frac{\D Y_{\Delta L\vphantom{()}}}{\D x}=-\frac{\D Y_{\Delta (B-L)}}{\D x}=Ax^{-2}.
\end{align}
These equations have analytic solutions. After the densities are frozen in, lead to $Y_f=Y_S=S$ and $Y_{\Delta (B-L)}=-A$. As the temperature drops below $m$, the asymmetric $f\rightarrow lHS$ decays double the abundance of scalar dark matter particles $S$ and reduce the asymmetry to $-\tfrac{3}{4}A$. The latter is converted into the baryon asymmetry via sphalerons as \cite{Kuzmin:1985mm, Harvey:1990qw}
\begin{align}\label{eq16}
Y_{\Delta B}=\frac{28}{79}Y_{\Delta(B-L)}=-\frac{21}{79}A.
\end{align}
Putting everything together, we obtain the ratio of the observed baryon and dark matter densities \cite{ParticleDataGroup:2024cfk} 
\begin{align}\label{eq17}
\frac{\Omega_{\text{b}}}{\Omega_{\text{dm}}}=\frac{Y_{\Delta B}}{Y_S}\frac{m_p}{m'}=\frac{63}{632}\frac{\theta}{\pi}\frac{m_p m_\nu}{v^2}\frac{m}{m'}\approx 0.19
\end{align}
where $\theta=-\Im[\lambda^2\lambda'^*]/(\vert\lambda\vert^2\vert\lambda'\vert$) is invariant under rescaling of all couplings, and we always have $\vert\theta\vert\leq 1$. Then, from Eq.~\eqref{eq17}, $m/m'\gtrsim10^{16}$. The hierarchy between the electro\-weak scale and neutrino masses thus results in a hierarchy between the heavy fermion mass $m$, determining the scale of the asymmetry generation, and the dark matter mass $m'$. Now, we can better understand why two hidden-sector particles were needed. For a single fermion $\chi$ interacting as in Eq.~\eqref{eq5}, asymmetric $lH\rightarrow 3\chi$ production yields a relation similar to Eq.~\eqref{eq17} with $m$ to $m'$ ratio replaced by unity. In such a case, nothing compensates for the smallness of the neutrino masses, and the asymmetry is suppressed by orders of magnitude.

\section{Boltzmann equations and numerical analysis for various benchmark scenarios}\label{sect4}

We proceed with a more general setup, including washout effects, although we still neglect  spectator processes \cite{Buchmuller:2001sr}. Using the classical Maxwell-Boltzmann distributions results in $Y^\eq_{H}=Y^\eq_l=45/(\pi^4 h_*)$ for the Higgs and lepton equilibrium densities, where $h_*=106.75$. The full set of the Boltzmann equations then read
\begin{align}
sHz\frac{\D Y_f}{\D z}=
&\bigg(1-\frac{Y_f}{Y^\eq_f}\frac{Y_S}{Y^\eq_S}\bigg)\gamma^\eq_{lH\rightarrow Sf} 
-\bigg(\frac{Y_f}{Y^\eq_f}-\frac{Y_S}{Y^\eq_S}\bigg)
\bigg(\gamma^\eq_{fH\rightarrow S\lbar}+\gamma^\eq_{fl\rightarrow S\Hbar \vphantom{\lbar}}\bigg)
\label{eq18}\\
&-\bigg(\frac{Y_f}{Y^\eq_f}-\frac{Y_S}{Y^\eq_S}\bigg)
\bigg(\gamma^\eq_{f\rightarrow lHS\vphantom{\lbar}}+\gamma^\eq_{f\rightarrow \lbar\Hbar S}\bigg)\nonumber\\
sHz\frac{\D Y_S}{\D z}=
&\bigg(1-\frac{Y_f}{Y^\eq_f}\frac{Y_S}{Y^\eq_S}\bigg)\gamma^\eq_{lH\rightarrow Sf} 
+\bigg(\frac{Y_f}{Y^\eq_f}-\frac{Y_S}{Y^\eq_S}\bigg)
\bigg(\gamma^\eq_{fH\rightarrow S\lbar}+\gamma^\eq_{fl\rightarrow S\Hbar \vphantom{\lbar}}\bigg)
\label{eq19}\\
&+\bigg(\frac{Y_f}{Y^\eq_f}-\frac{Y_S}{Y^\eq_S}\bigg)
\bigg(\gamma^\eq_{f\rightarrow lHS\vphantom{\lbar}}+\gamma^\eq_{f\rightarrow \lbar\Hbar S}\bigg)\nonumber\\
sHz\frac{\D Y_{\Delta L}}{\D z}=
&\bigg(1-\frac{Y_f}{Y^\eq_f}\frac{Y_S}{Y^\eq_S}\bigg)\Delta\gamma^\eq_{lH\rightarrow Sf}-\frac{4}{3}\frac{Y_{\Delta L}}{Y^\eq_l}\bigg(
\gamma^\eq_{Sf\rightarrow lH\vphantom{\lbar}}
+2\gamma^\eq_{\lbar\Hbar \rightarrow lH} 
+2\gamma^\eq_{\Hbar \Hbar \rightarrow ll\vphantom{\lbar}}
\bigg)\label{eq20}\\
&-\frac{Y_{\Delta L}}{Y^\eq_l}\bigg(\frac{Y_f}{Y^\eq_f}+\frac{1}{3}\frac{Y_S}{Y^\eq_S}\bigg) \gamma^\eq_{f\Hbar \rightarrow Sl}-\frac{Y_{\Delta L}}{Y^\eq_l}\bigg(\frac{Y_S}{Y^\eq_S}+\frac{1}{3}\frac{Y_f}{Y^\eq_f}\bigg)
\gamma^\eq_{S\Hbar \rightarrow fl}\nonumber\\
&+\bigg(\frac{Y_f}{Y^\eq_f}-\frac{Y_S}{Y^\eq_S}\bigg)\Delta\gamma^\eq_{f\rightarrow lHS}
-\frac{4}{3}\frac{Y_{\Delta L}}{Y^\eq_l}\frac{Y_S}{Y^\eq_S}\gamma^\eq_{f\rightarrow lHS\vphantom{\lbar}}
\nonumber
\end{align}
where $z=m/T$. In Eq.~\eqref{eq20}, we approximate the asymmetries of specific flavours by $Y_{\Delta L}/3$, and, from the symmetry of the model in Eq.~\eqref{eq6}, we also employ $Y_{\Delta H}=Y_{\Delta L}$. Explicit analytical expressions for the reaction rates entering Eqs.~\eqref{eq18}-\eqref{eq20} are listed in \ref{app1}. 

\begin{figure*}[t]
\subfloat{\label{fig3a}}
\subfloat{\label{fig3b}}
\subfloat{\label{fig3c}}
\subfloat{\label{fig3d}}
\centering\includegraphics[scale=1]{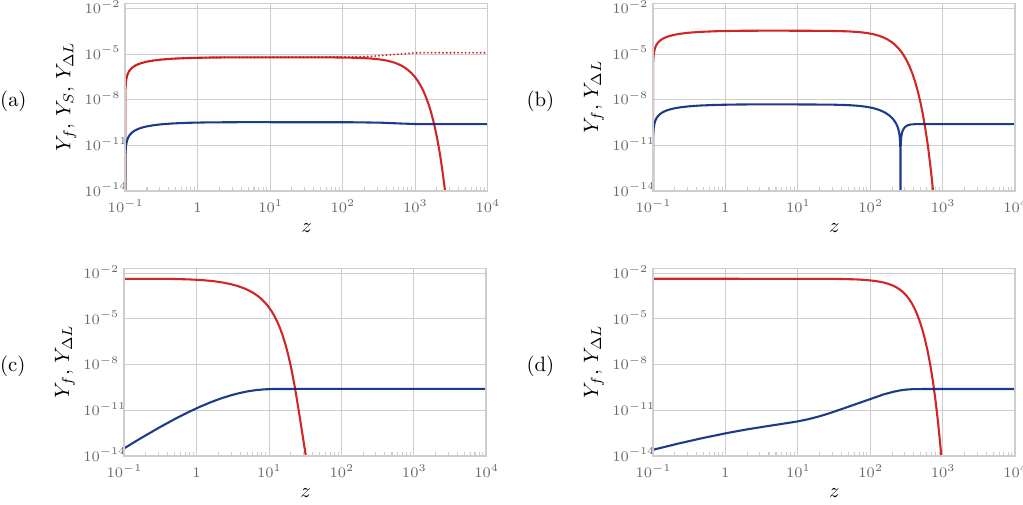}
{\caption{\label{fig3} 
The evolution of $Y_f$ (solid red line), $Y_S$ (dotted red line in {\bf a}), and $\vert Y_{\Delta L}\vert$ (solid blue line). Here {\bf a} corresponds to the freeze-in scenario with $\Lambda=10^{14}~\text{GeV}$, $\TR=5\times 10^{12}~\text{GeV}$, $m=5\times 10^{11}~\text{GeV}$, $\vert\lambda\vert^2=1.8\times 10^{-3}$, and $\theta=0.8$. In {\bf b} , $Y^{\vphantom{\eq}}_S=Y^\eq_S$, while the parameters were the same except for $\vert\lambda\vert^2=7\times 10^{-3}$. Both {\bf c}  and {\bf d}  show the freeze-out case with $\Lambda=10^{12}~\text{GeV}$ and $m=10^{10}~\text{GeV}$. In {\bf c} , $\vert\lambda\vert^2=0.1$ and $\theta=0.25$, while in {\bf d}  we used $\vert\lambda\vert^2=10^{-4}$ and $\theta=0.12$.}}
\end{figure*}

Interestingly enough, within the minimal model defined in Eq.~\eqref{eq2}, the matter-antimatter asymmetry and dark matter relic density can be obtained in a variety of scenarios depending on the initial conditions. Numerical solutions of Eqs.~\eqref{eq18}-\eqref{eq20} are plotted in Fig.~\ref{fig3}, where the parameters were tuned to reproduce the observed values. First, in Fig.~\ref{fig3a}, we show the freeze-in of the $f$ and $S$ densities while the asymmetry is produced. At $z\simeq 10^3$, the asymmetric $f\rightarrow lHS$ decays slightly reduce its final value. Remarkably, it only differs from what can be obtained from Eq.~\eqref{eq16} by roughly $8\%$. The dark matter relic density matches the observed value for $m'\simeq 40~\text{keV}$.

The situation in Fig.~\ref{fig3b} is different. Initially, the scalars $S$ were in thermal equilibrium with the Standard Model. When the temperature dropped below $m'$, their comoving density froze out through the Higgs portal \cite{Silveira:1985rk, Burgess:2000yq} or their own self-interactions \cite{Bernal:2015xba}. This occurred much later than the part of the story shown in Fig.~\ref{fig3b}, where the $Y^{\vphantom{\eq}}_S=Y^\eq_S$ constant is not shown. We instead focus on the freeze-in of fermions $f$. For temperatures below $m$, their decays became too slow to follow the equilibrium distribution, and $Y^{\vphantom{\eq}}_f$ was much larger than $Y^\eq_f$. The scattering contribution to the asymmetry was dominated by the second term in the first round bracket in Eq.~\eqref{eq20} and the sign of the asymmetry changed at $z\simeq 250$. Note that in this case, the dark matter mass $m'$ becomes independent from the fermion mass $m$.

Finally, in Figs.~\ref{fig3c} and \ref{fig3d}, both $f$ and $S$ were initially thermalized. We used a larger value of $\vert\lambda\vert$ in Fig.~\ref{fig3c} and the asymmetry production was dominated by $lH\rightarrow fS$ reactions. In Fig.~\ref{fig3d}, $\vert\lambda\vert$ was smaller, and $Y_f$ decreased more slowly, producing larger departures from equilibrium that, similarly to the sign flip in Fig.~\ref{fig3b}, led to the asymmetry enhancement at $z\simeq 20-300$.

\section{Summary and discussion}\label{sect5}

Starting from general assumptions, we have constructed a minimal model for leptogenesis, dark matter, and neutrino masses based on two $d=5$ effective operators: the Weinberg operator and a portal operator for the dark-sector interactions with the Standard Model. Two new particles, with a substantial mass hierarchy, were needed. A heavy Majorana fermion played an essential role in the lepton-number asymmetry generation through the asymmetric production or decay processes. A light neutral scalar particle was proposed as a dark matter candidate. After a simplified analytical treatment of the freeze-in scenario, a full set of the Boltzmann equations was solved numerically with various initial conditions. The observed values of the baryon asymmetry and dark matter relic density have been reproduced.

In our analysis, we were agnostic about the underlying UV-complete theory. However, several approaches leading to a similar effective description were considered in the literature. In Ref.~\cite{Bento:2001rc}, off-shell Majorana fermions were used to mediate the interactions of the Standard-Model leptons with the mirrored dark sector. For the asymmetry generation, however, the authors effectively employed an additional $d=6$ operator that, with our notation, would be analogous to $S\Bar{f}\slashed{\partial}P_L lH$. This would lead to asymmetries scaling as $T^8/\Lambda^4$, instead of $mT^6/\Lambda^3$ as in Eq.~\eqref{eq11}.\footnote{See the discussion above Eq.~(7a) in Ref.~\cite{Bento:2001rc}.} We instead rely on a more minimal case of the single $d=5$ portal operator and a non-zero mass of fermions $f$ entering the asymmetry calculation.

In Refs.~\cite{Chun:2011cc, Falkowski:2011xh, Falkowski:2017uya, Barman:2022scg, Herrero-Garcia:2024tyh}, the neutrino-portal interactions were considered with the asymmetry sourced by right-handed neutrino decays. Notably, the authors of Ref.~\cite{Herrero-Garcia:2024tyh} recently investigated a minimal scenario incorporating dark matter, leptogenesis, and light neutrino masses. In this work, we argue that the setup can be made even more minimal by integrating all right-handed neutrinos out. Unlike in Ref.~\cite{Herrero-Garcia:2024tyh}, it is essential for us that the fermionic dark-sector particles are heavy, while scalars become the stable dark matter candidates.

\begin{acknowledgements}
The authors thank Fedor \v{S}imkovic and Eli\v{s}ka Eckerov\'{a} for a valuable discussion regarding the latest neutrino mass limits. T. Bla\v{z}ek and P. Mat\'{a}k were supported by the Slovak Grant Agency VEGA, project No. 1/0719/23, and the Slovak Education Ministry Contract No. 0466/2022. 
\end{acknowledgements}

\appendix
\section{Reaction rates and asymmetries}\label{app1}

In Eqs.~\eqref{eq18}-\eqref{eq20}, the reaction rates and their asymmetries were treated as functions of $z=m/T$. We obtain for the symmetric rates
\begin{align}
&\gamma^\eq_{lH\rightarrow Sf\vphantom{\lbar}}=\gamma^\eq_{lS\rightarrow\Hbar f\vphantom{\lbar}}=\frac{\vert\lambda\vert^2}{128\pi^5}\frac{m^6}{\Lambda^2}\frac{K_1(z)+K_3(z)}{z^3}\\
&\gamma^\eq_{fl\rightarrow S\Hbar \vphantom{\lbar}}=\frac{\vert\lambda\vert^2}{16\pi^5}\frac{m^6}{\Lambda^2}\frac{K_2(z)}{z^4}\\
&\gamma^\eq_{f\rightarrow lHS}=\frac{\vert\lambda\vert^2}{768\pi^5}\frac{m^6}{\Lambda^2}\frac{K_1(z)}{z}
\end{align}
where $K_i(z)$ denotes the modified Bessel functions of the second kind. The asymmetries are calculated using Eq.~\eqref{eq2} and result in 
\begin{align}
&\Delta\gamma^\eq_{lH\rightarrow Sf\vphantom{\lbar}}=
\frac{3\Im[\lambda^2\lambda'^*]}{256\pi^6}\frac{m^7}{\Lambda^3}\frac{K_3(z)}{z^3}\\
&\Delta\gamma^\eq_{f\rightarrow lHS}=
-\frac{\Im[\lambda^2\lambda'^*]}{2048\pi^6}\frac{m^7}{\Lambda^3}\frac{K_1(z)}{z}
\end{align}
while asymmetries of other reactions can be expressed using the $CPT$ and unitarity constraints, as in Eqs.~\eqref{eq9} and \eqref{eq10}.

\bibliographystyle{apsrev4-1.bst}
\bibliography{CLANOK.bib}

\end{document}